# Bridging atomistic simulations and thermodynamic hydration models of aqueous electrolyte solutions

Xiangwen Wang,[†] Simon L. Clegg,[*,‡] and Devis Di Tommaso,[*,†]

[†] Department of Chemistry, Queen Mary University of London, Mile End Road, London, E1 4NS, United Kingdom

[‡] School of Environmental Sciences, University of East Anglia, Norwich, NR4 7TJ, United Kingdom

**ABSTRACT:** Chemical thermodynamic models of solvent and solute activities predict the equilibrium behaviour of aqueous solutions. However, these models are semi-empirical. They represent micro-scale ion and solvent behaviours that control the macroscopic properties using small numbers of parameters whose values are obtained by fitting to activities and other partial derivatives of the Gibbs energy measured for the bulk solutions. We have conducted atomistic simulations of aqueous electrolyte solutions ($MgCl_2$ and $CaCl_2$) to determine the parameters of aqueous thermodynamic hydration models. We have implemented a cooperative hydration model to categorize the water molecules in electrolyte solutions into different subpopulations. The value of the electrolyte-specific parameter, *k*, was determined from the ion-affected subpopulation with the lowest absolute value of the free energy of removing the water molecule. The other equilibrium constant parameter, $K_1$, associated with the first degree of hydration, was computed from the free energy of hydration of hydrated clusters in solution. The hydration number, *h*, was determined from a reorientation dynamic analysis of the water subpopulations compared to bulk-like behaviour. The computed values of these parameters were inserted in the Stokes & Robinson (*Journal of Solution Chemistry* **1973**, *2*, 173-191) and Balomenos (*Fluid Phase Equilibria* **2006**, *243*, 29-37) models, which were applied to evaluate the osmotic coefficients of $MgCl_2$ solutions. Such an approach removes the dependence on the availability of experimental data and could lead to aqueous thermodynamic models capable of estimating the values of solute and solvent activities, thermal and volumetric properties for a wide range of composition and concentrations.

## 1. INTRODUCTION

Aqueous thermodynamic models of solvent and solute activities are used to predict the equilibrium behaviour and chemical speciation of natural aqueous solutions (oceans, brines, groundwaters, atmospheric aerosols) and industrial systems (ionic liquids, underground contaminants, fluids used for oil and gas processing).[2-6] Some of the most widely used thermodynamic models to estimate the activity coefficients of electrolyte solutions are semi-empirical.[5] They attempt to represent key micro-scale ion and solvent behaviours that control the macroscopic properties in a simplified way using small numbers of parameters whose values are obtained by fitting to activities and other partial derivatives of the Gibbs energy that have been measured for the bulk solutions. Examples include the Pitzer [7] and Pitzer-Simonson-Clegg models,[8, 9] which are applied in aqueous geochemistry and atmospheric science, and the non-random two-liquid (NRTL) models of Chen,[10, 11] used to describe mixed solvent systems of industrial solutions. However, key molecular-level behaviours such as hydration and ion pairing are generally not treated explicitly in these semi-empirical models and when extrapolated beyond the range of concentration to which they were fitted, they tend to give inaccurate thermodynamic properties.[5] In contrast to these approaches, some thermodynamic hydration models attempt to represent the microphysical hydration processes more directly within their theoretical framework (*vide infra*) through the incorporation of hydration (formation of solvent shells around ions).[12] Although their fundamental assumption (the ion-hydration is the principal determinant of activity) is different from that of the ion interaction and local composition models, they are also highly simplified and contain empirical fitted parameters.[17] Properties such as water activity and freezing point depression of simple electrolyte solutions have been shown to correlate very simply with concentration to calculated a range of hydration number, which gives some encouragement to the use of these approaches.[47]

Explicit-water molecular dynamics (MD) simulations and static density functional theory (DFT) calculations, with a dielectric continuum description of the solution environment, can provide an atomistic view of the processes of ion hydration and ion association in solution. These atomistic simulation methods have been extensively applied to quantify the energetics of ion-water and ion-ion interaction and to compute single-particle, pair, and collective dynamical properties of aqueous electrolyte solutions.[13-16] Therefore, advances in the field may be achieved if the parameters contained in thermodynamic hydration models could be obtained from atomistic simulations of aqueous electrolyte solutions rather than experimental data. Such an approach, if successful, would remove the dependence from the availability of experimental data and pave the way to develop aqueous thermodynamic models capable of estimating the values of solute and solvent activities, thermal and volumetric properties for a wide range of composition and concentrations.

This study aims to determine the link between the parameters controlling aqueous thermodynamic hydration models and the molecular-level processes occurring in electrolyte solutions. We have conducted MD simulations and DFT calculations of hydrated alkaline earth metal ions ($Mg^{2+}$ and $Ca^{2+}$) and aqueous electrolyte solution ($MgCl_2$ and $CaCl_2$) with different concentrations, with the results of the simulations being used to determine the parameters of thermodynamic hydration models: the hydration number $h$; the hydration stages $n$; the chemical equilibrium constant parameters $K$ and $k$. The values of the parameters computed from this first principle approach were then used directly in the hydration models of Stokes & Robinson[17] and Balomenos,[18] which were then applied to compute osmotic coefficients of aqueous $MgCl_2$ solution.

### 1.1 Overview of thermodynamic stepwise hydration model

The stepwise hydration-equilibrium models introduced by Stokes and Robinson (S&R),[17] and later improved by Schönert,[19, 20] assume that a cation can possess various discrete degrees of hydration related by stepwise hydration equilibria. Cations ($C$) become hydrated through a stepwise process where a single water molecule binds to the ion-water cluster during each degree of hydration:

$$C \cdot (H_2O)_{i-1} + H_2O \rightleftharpoons C \cdot (H_2O)_i \qquad (1)$$

In Eq. 1, each hydration step $i$ is controlled by an equilibrium constant relating the activities of the hydrated species and water:

$$K_i = \frac{a_i}{a_{i-1} a_W} \qquad (2)$$

The competition between neighbouring solvent molecules in the coordinate shells will reduce the strength of binding from one hydration step to the next. For each successive step, the assumption is that the standard free energy change for attachment of a water molecule decreases by a constant amount $RT\ln(k)$, where $k$ is an electrolyte-specific parameter. This simplified assumption works in R&S's electrolyte solution model but may not be realistic. It may also be that the model can represent activity data using equilibrium constants $K$ that are related to each other in different ways. In the R&S approach, the equilibrium constants for each step are related by:

$$\begin{aligned} K_1 &= K \\ K_2 &= kK \\ &\dots \\ K_i &= k^{i-1}K \\ &\dots \\ K_n &= k^{n-1}K \end{aligned} \qquad (3)$$

where $i$ is the number of sites of hydration and goes from 1 to $n$. If $c_i$ is the stoichiometric molar concentration of the $i$-th hydrated species, then the average hydration number $h$ is then defined as:

$$h = \frac{\sum i c_i}{\sum c_i} \qquad (4)$$

In the S&R hydration model,[1] the partial molar volume of $i$-th hydrated species, $\bar{V}_i$, is considered to be linearly dependent on the degree of hydration:

$$\bar{V}_i = \bar{V}_0 + i\bar{V}_W \qquad (5)$$

where $\bar{V}_0$ is the partial molar volume of zero hydration degree, and $\bar{V}_W$ is the partial molar volume of a water molecule. The value of the maximum number of hydration sites, $n$, corresponds to the hydration number, $h$. In the S&R model, the water activity and the mean osmotic coefficient depend on the equilibrium constant, the hydration number, and the partial molar volume of different hydration species:[21]

$$\ln(a_A) = \ln(1 - c\bar{V}_h) + c(\bar{V}_h - \nu\bar{V}_W) + \frac{\kappa^3}{24\pi N_A} \cdot \bar{V}_W S(\kappa a) \quad (6)$$

where $a$ is the distance of closest approach in the Debye-Hückel theory, $\nu$ is the stoichiometric number of the solute, $N_A$ is the Avogadro's constant, $\kappa$ is Debye–Hückel reciprocal length expressed in the SI unit system and the molar concentration scale, and the function $S(x)$ comes from the Debye theory:

$$S(x) = \left(\frac{3}{x^3}\right)\left[1 + x - \frac{1}{1+x} - 2\ln(1+x)\right] \quad (7)$$

For a single salt solute with molal activity, the osmotic coefficient can be then written as:

$$\phi = -\frac{\ln(a_A)}{\nu m M_A} \quad (8)$$

Schönert's model[19] was developed based upon the S&R approach by considering the hydration of both cation and anion. The anion hydration is assumed to take place with the same stepwise procedure as the cation.[19, 20] The hydration equilibrium constant of each step is given by the following expression:

$$K_{p,i} = (k_p)^i \binom{n_p}{i} \quad (9)$$

In Eq. 9, $k_p$ is the binding hydration constant, where $p$ refers to $c$, the cation, or $a$, the anion. The hydration equilibrium constant $K_{p,i}$ for a hydrated ion will then depend on the distinct combinations of the identical $i$ water molecules on the available $n_p$ identical hydration sites. Following Schönert's assumption that the hydrated species form a semi-ideal mixture with the solvent, the average hydration number for each ion is related to the mole fraction of the bulk solvent water $x_W$:

$$h_p = \frac{\sum iC_{p,i}}{\sum C_{p,i}} = \frac{n_p k_p x_W}{1 + k_p x_W} \quad (10)$$

The hydration model of Balomenos combines the Pitzer–Debye–Hückel theory,[22] which used statistical thermodynamics to retain a more realistic ionic distribution, removes the charging process and incorporates a hardcore repulsion factor, with the stepwise hydration model of Schönert.[18] The long-range ionic interaction is described by Pitzer's hardcore repulsion factor,[22] and hydration association is represented by Schönert's two electrolyte parameters hydration sites and the binding hydration constant.[19, 20] In the Balomenos' model the osmotic coefficient is defined as:

$$\phi = -\frac{\ln(x_W)}{1000 V_W c} - \frac{\kappa^3}{6Dc(1+\kappa\bar{a})} + \frac{\pi\bar{a}c}{3}\left(2\bar{a}^2 + \frac{\kappa^4}{\left(Dc(1+\kappa\bar{a})\right)^2}\right) \quad (11)$$

In Eq. 11, $c$ is the total molar concentration of all ions, $D = 4\pi 1000 N_A$, $\kappa$ is Debye–Hückel reciprocal length, and $\bar{a}$ is the average distance of closest approach, which is defined as:

$$\frac{\bar{a}}{2} = r_{CA}^h = [0.5((r_C)^3 + (r_A)^3) + h(r_W)^3]^{1/3} \quad (12)$$

where $r_{CA}^h$ is the radius of the average ionic hydration sphere, $r_C$ is the radius of cation, $r_A$ is the radius of anion, $r_W$ is the radius of water and $h$ is the maximum number of hydration sites (here the hydration number). Balomenos' model uses only one electrolyte parameter, the maximum number of hydration sites, which includes contributions from ion-specific interactions between a single ion and solvent molecules, and electrolyte specific interactions between multiple ions and solvent molecules.

## 2. COMPUTATIONAL DETAILS

### 2.1 Classic molecular dynamics simulation

We used classical MD simulations of magnesium chloride solutions, $MgCl_2$(aq), and calcium chloride solutions, $CaCl_2$ (aq), with concentrations ranging from 0.1 to 2.8 mol.kg$^{-1}$. These simulations have been used to characterize the structure and low-frequency dynamics of water and estimate the hydration number, $h$, of this electrolyte as a function of concentration. Furthermore, free-energy difference calculations from ensembles generated by MD simulation have also been conducted to evaluate the electrolyte-specific parameter, $k$. All MD simulations were carried out using GROMACS (v. 2019.4).[23] The models of aqueous electrolyte solution were based on the full atomistic treatment of the solute and solvent molecules. Water molecules were described with the three-site SPC/E water model,[24] while "scaled charge" Empirical Continuum Correction (ECC) force fields were used to describe the ion-water and ion-ion interaction.[25-27] In this approach, the fast electronic polarization is taken into account in a mean field approach and implemented numerically by scaling the charges of the ions. Here, we have used the Lennard-Jones potentials for $MgCl_2$(aq) and $CaCl_2$(aq) parameterized by Duboue-Dijon et al.,[27] in which the charges for the cations (magnesium and calcium) and chlorine ions are set to +1.7 and –0.85, respectively.

The MD simulations were conducted using the following computational protocol. A cubic box was filled with approximately 800 water molecules. The MD simulations were conducted in the isothermal-isobaric ensemble (NPT, P = 1 atm and T = 300K) for 1 ns to generate an equilibrated aqueous solution. The last configuration was then used to generate $MgCl_2$(aq) and $CaCl_2$(aq), with concentrations ranging from 0.1 to 2.8 mol kg$^{-1}$, by randomly replacing $N$ water molecules with $N/3$ $Mg^{2+}$ and $2N/3$ $Cl^-$ ions and by making sure that the initial configuration did not contain contact ion pairs. The volume of the simulation box was further minimized in the NPT ensemble followed by a production period of 1 ns. Configurations were saved every 1 ps. Long-range electrostatic interactions were accounted for using the Particle Mesh Ewald (PME) method with a 12 Å cutoff for nonbonded interactions. Simulations in the NPT ensemble used the Velocity-rescale thermostat[28] and Parrinello−Rahman pressure coupling methods[29] with a 2 ps time constant. Bonds involving hydrogen atoms were constrained by the LINCS algorithm[30].

The free energy calculations of removing a specific water molecule in the electrolyte solution were conducted with the GROMACS 'bar' module which implements the Bennett Acceptance Ratio (BAR) method combined with the PLUMED plugin (v. 2.4.1).[31, 32] To avoid the selected water molecule moving between different coordination shells during the MD simulation, we imposed two harmonic biases potentials, both with a constant of 1000 kJ mol$^{-1}$, along the reaction coordinates defined by the distances between the cation and the water, and between the anion and the water. MD trajectories were visualised with the Visual Molecular Dynamics (VMD)[33] and Chimera software.[34]

### 2.2 DFT calculations

Electronic structure calculations were carried out with the Gaussian09 code.[35, 36] Geometry optimization and frequency calculations were both conducted with the long-range corrected hybrid density functional with damped atom-atom dispersion corrections (B97X-D)[37] together with the Pople basis set 6-311++G(d,p) basis set. The solvation models used to conduct the solution phase calculation were the conductor-like polarizable continuum model (CPCM) and the polarizable continuum model using the integral equation formalism variant (PCM).[38] The free energies of the stepwise process (Eq. 1) where a single water molecule binds to the ion-water cluster were computed according to the following equation:

$$\Delta G^* = G^*_{C^+(H_2O)_i} - G^*_{C^+(H_2O)_{i-1}} - G^*_{H_2O} \qquad (14)$$

where $G^*_X$ is the total Gibbs free energy of species ($X = C^+(H_2O)_i$, $C^+(H_2O)_{i-1}$, or $H_2O$) in the liquid phase at 300K. Following the recommendation by Ho et al. that free energies of molecules in solution should be obtained from separate gas- and solution-phase calculations,[39] the values of $G^*_X$ were calculated using the following expression[40]:

$$G^*_X = E_{e,gas} + \delta G^{\circ}_{VRT,gas} + \Delta G^*_{solv} + RT\ln[\tilde{R}T] \qquad (15)$$

where $E_{e,gas}$ is the gas-phase total electronic energy of the gas-phase optimized geometry of the species $X$, $\delta G^{\circ}_{VRT,gas}$ is the vibrational-rotational-translational contribution to the gas-phase Gibbs free energy at 300K under a standard-state partial pressure of 1 atm, $\Delta G^*_{solv}$ is the solvation free energy of the solute corresponding to transfer from an ideal gas at a concentration of 1 mol dm$^{-3}$ to an ideal solution at a liquid-phase concentration of 1 M, and the $RTln[\tilde{R}T]$ term is the free energy change of 1 mol of an ideal gas from 1 atm to 1 mol dm$^{-3}$ ($RT\ln[\tilde{R}T]$ = 1.89 kcal mol$^{-1}$ at 300K, where $\tilde{R}$ = 0.082052 K$^{-1}$).[41, 42]

## 3. RESULTS AND DISCUSSION

### 3.1 Subpopulation water classification method

In an aqueous electrolyte solution, a water molecule of the first- or second-coordination shells of the cation/anion, was shared by both ions, part of the bulk. We have implemented a cooperative hydration model to categorize the water molecules in electrolyte solutions into different subpopulations (**Figure. 1**).[53] Each water molecule in the solution is labelled $W_{ab}$, where $a, b = 1, 2, B$, depending on the position of oxygen and hydrogen atoms from the cation ($Mg^{2+}$ or $Ca^{2+}$) and anion ($Cl^-$). The subscript in $W_{ab}$ is set to 1 when the oxygen or hydrogen atom is in the first coordination shell of the nearest ion, to 2 when the oxygen or hydrogen atom is in the second coordination shell of the nearest ion, and to B when it is beyond the second shell. Assignments were made by comparing the distance between oxygen and the nearest cation within the first and second magnesium coordinate shells, and the distance between hydrogen and the nearest chloride within the first and second chloride coordinate shells (any one of the two hydrogens in the water molecule). We have assumed that water molecules in the same subpopulation have the same behaviour.

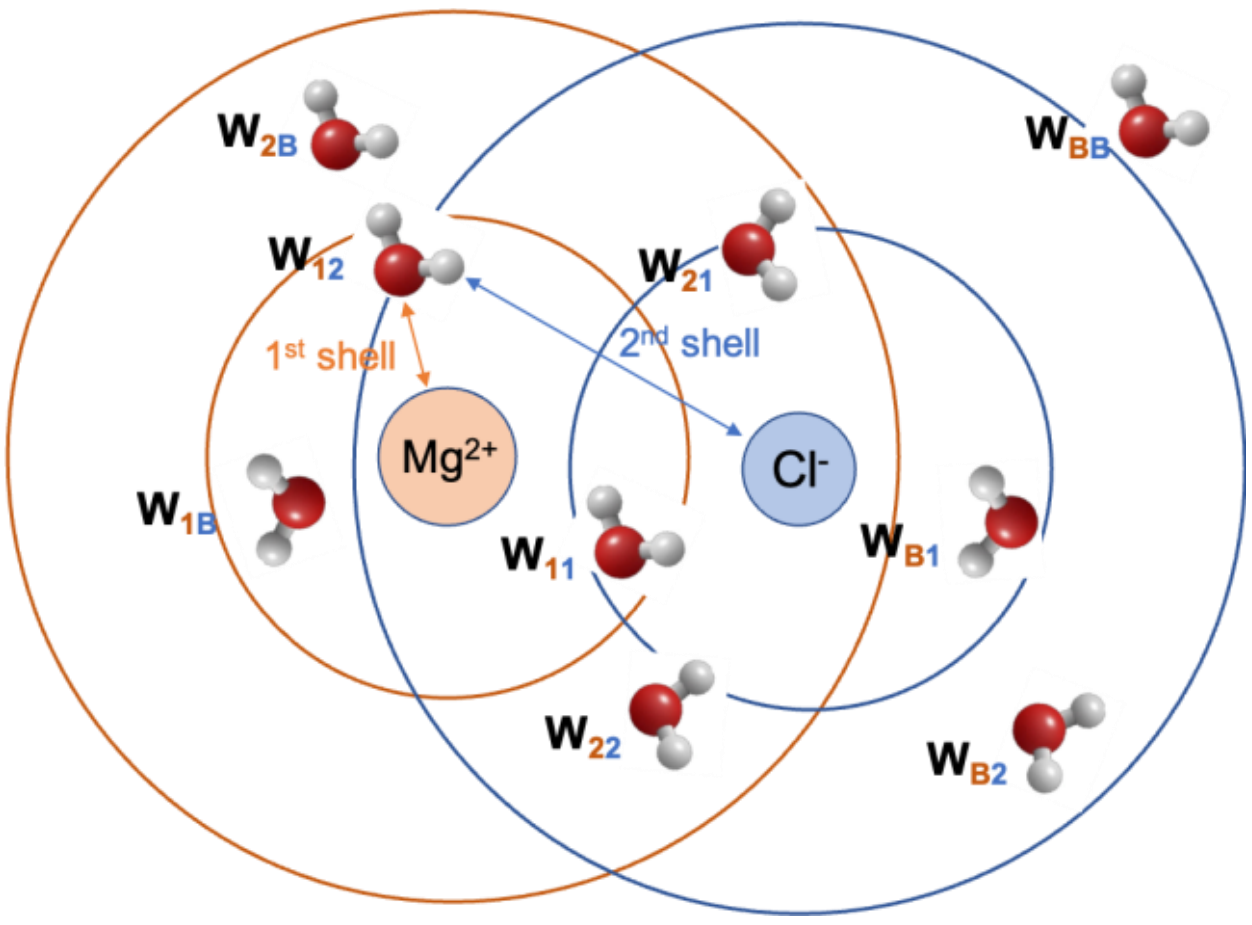

**Figure 1.** Definition of water subpopulations for a solvent separated $Mg^{2+}$ and $Cl^-$ ion pair showing the categorization of one $H_2O$ molecule in the $W_{12}$ subpopulation: O atom in the 1st coordination shell of $Mg^{2+}$; both H atoms in the 2nd coordination shell of $Cl^-$.

### 3.2 Equilibrium constant *k* and *K* parameters

The stepwise hydration process to form the fully hydrated cation $[C(H_2O)_h]^{z+}$ is given by the following equation:

$$[C(H_2O)_{h-1}]^{z+} + H_2O \rightleftharpoons [C(H_2O)_h]^{z+} \qquad -\Delta G_n,\ K_n = k^{n-1}K \qquad (16)$$

where $h$ is the hydration number, which represents the maximum number of water molecules influenced by the cation, $n$ is the maximum degree of hydration, and $K_n$ is the equilibrium constant of the stepwise process that is related to the free energy of removing the $h$-th water molecule from $C(H_2O)_h$ by $\Delta G_n = RT\ln K_n$. The hydration of the fully hydrated cation

$$[C(H_2O)_h]^{z+} + H_2O \rightleftharpoons [C(H_2O)_{h+1}]^{z+} \qquad -\Delta G_s,\ K_{n+1} = k^n K = K_W \tag{17}$$

will have an equilibrium constant equivalent to the hydration of a water molecule in pure bulk water ($K_W$), which is related to the free energy needed for removing one bulklike water from the fully hydrated cluster, $\Delta G_s = RT\ln(K_s)$. Therefore, the electrolyte-specific parameter $k$, which is the standard free energy change for successive step water attachment in the S&R hydration model, $RT\ln(k)$, is given by:

$$k = \frac{K_W}{K_n} \tag{18}$$

Based on the assumption that water molecules in the same subpopulation have equal ability to bind the ion-water cluster, the electrolyte-specific parameter $k$ can be determined from the ion-affected subpopulation with the lowest absolute value of $\Delta G_n$:

$$-RT\ln(k) = min|\Delta\Delta G| = |\Delta G_n - \Delta G_s| \tag{19}$$

The value of $\Delta G_n$ for each water subpopulation in an aqueous electrolyte have been computed using a computational protocol based on the BAR method, as detailed in 2.1. This methodology has been used to determine the $k$ parameter of aqueous $MgCl_2$ solution for concentrations ranging from 0.1 to 2.8 mol kg$^{-1}$ (**Figure. 2**). The average values of the free energy $\Delta G_n$ for the subpopulation of $MgCl_2$(aq) is reported in **Figure. 2A**. To separate the effect of $Mg^{2+}$ and $Cl^-$ on the water molecules, we generated two standard lines. The orange dash line ($W_{BB}$ subpopulation) represents the free energy change of removing a bulklike water molecule. We use this line to compare the free energy values of the water molecules in the subpopulations without the effect of chloride: $W_{1x}$, $W_{22}$, $W_{2B}$, and $W_{B2}$. The sky-blue dash line represents the free energy change of removing a water molecule in the first coordinate shell of $Cl^-$ obtained from MD simulations of one isolated chloride in the pure water system. We use this line to compare the free energy change of water molecules belonging to subpopulations that could be affected by the chloride only: $W_{21}$, $W_{B1}$.

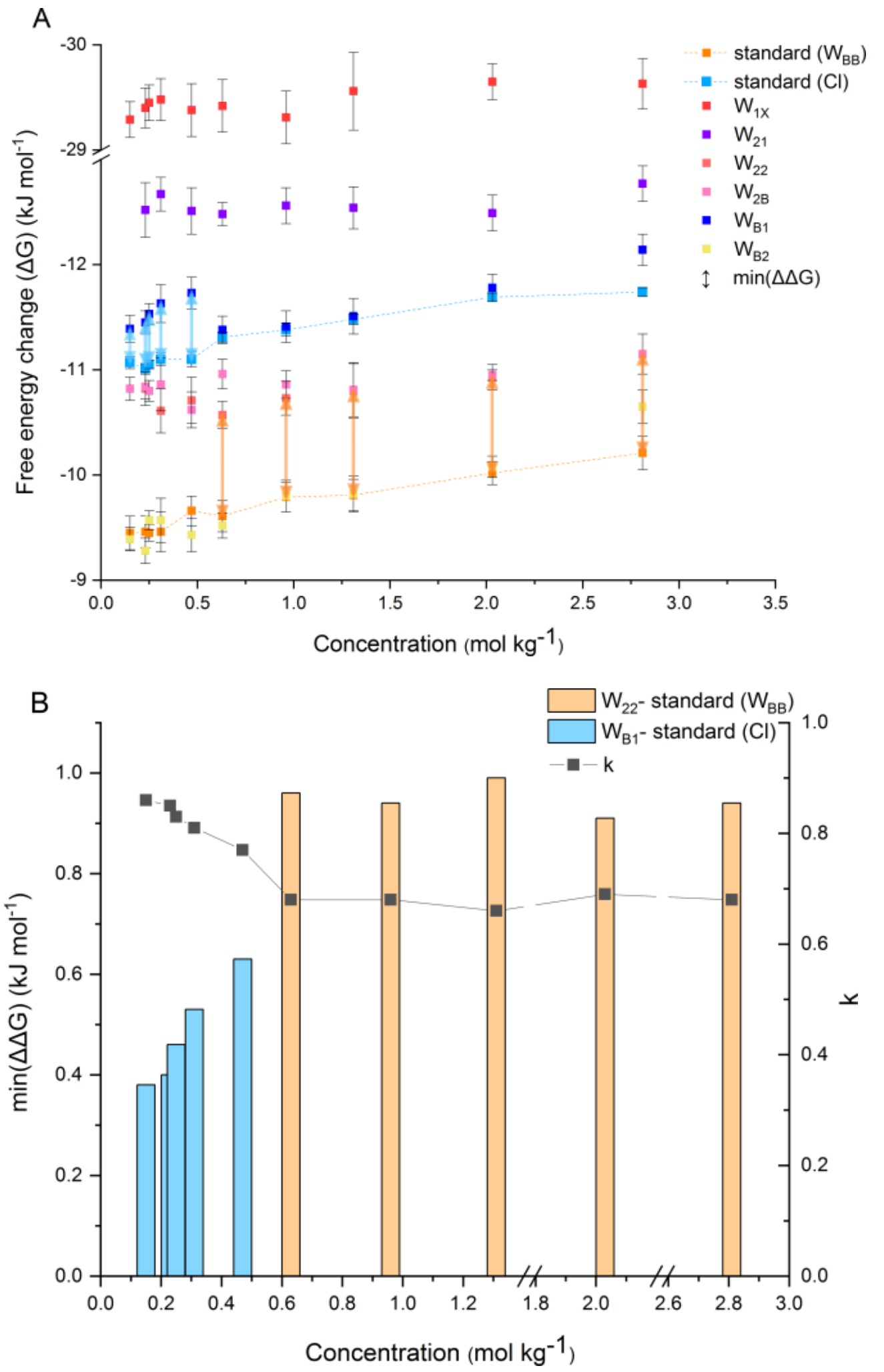

**Figure 2.** Results from free energy calculations of $MgCl_2$(aq) with concentrations from 0.1 to 2.8 mol kg$^{-1}$. **(A)** Average free energy change of removing a water molecule from each subpopulation. The orange dashed line ($W_{BB}$) is the free energy change of removing a bulk-like water molecule. The sky-blue dashed line is the free energy change of removing a water molecule in the first coordinate shell of $Cl^-$ obtained from simulations of an isolated ion in water. **(B)** Values of $min|\Delta\Delta G|$ and $k$ as a function of concentration. The blue bars are $min|\Delta\Delta G|$ values obtained from the difference of $\Delta G_n$ from the chloride-effected subpopulation ($W_{B1}$) and $\Delta G_s$ from the sky-blue dash line. The orange bars are $\Delta\Delta G$ obtained from the difference between $\Delta G_n$ from the Mg-effected subpopulations ($W_{22}$) and $\Delta G_s$ from the $W_{BB}$ subpopulation.

The variation of $min|\Delta\Delta G|$ of electrolyte solutions as a function of concentration is summarised in **Figure. 2B**, where the blue- and orange-coloured bar charts represent the chloride- and magnesium-affected water subpopulations, respectively. For molality lower than 0.6 mol kg$^{-1}$ the value of $min|\Delta\Delta G|$ comes from the $W_{B1}$ subpopulation, which corresponds to water molecules beyond the second coordinate shell of the cation but in the first coordinate shell of the anion. These water molecules can be considered non-bulklike. The values of the $k$ parameter are between 0.8 and 1. For more concentrated solutions, $min|\Delta\Delta G|$ is due to the $W_{22}$ subpopulation and the equilibrium constant parameter $k$ shows a constant trend around 0.7. In comparison, the value of k obtained from the application of the S&R model to $MgCl_2$(aq) (fitting code with the e04fyf routine from NAG library) is 0.9 (Supporting Information, Table S1).

The other equilibrium constant parameter, $K_1$, is associated with the first degree of hydration ($n$=1). $Mg^{2+}$ is a strongly hydrated ion with six water molecules stably coordinated to the ion.[43] Consequently, these water molecules will not participate in the stepwise hydration process. The DFT calculations were conducted to determine the free energies of the formation of the pseudo-first hydration process:

$$Mg^{2+}(H_2O)_6 + H_2O \rightleftharpoons Mg^{2+}(H_2O)_7 \qquad K_1 = K \tag{18}$$

with the seventh water molecule in the second coordinate shell binding onto the magnesium-water cluster. We calculated $G_X^*$, the total Gibbs free energy of species (X = $Mg^{2+}(H_2O)_6$, $Mg^{2+}(H_2O)_7$, or $H_2O$), and further got the Gibbs free energy change of the initial hydration process (Supporting Information, Table S2). According to the equation $\Delta G = -RTlnk$, the equilibrium constant parameter $K$ of $MgCl_2$ is 2.72.

### 3.3 Hydration numbers and hydration stages

We present the determination of the hydration number $h$ of an electrolyte solution using two methods based on the calculation of the following water-specific properties: the free energy of water removal; the water dipole reorientation dynamics.

#### *3.3.1 Free energy-based method*

The first method uses the equilibrium constant parameters determined from the free energies of the stepwise hydration process $[C(H_2O)_{h-1}]^{z+} + H_2O \rightleftharpoons [C(H_2O)_h]^{z+}$. By inserting the values $k$ and $K$ determined in Section 3.2 into Eq. 3, we can obtain the equilibrium constant $K_i$ ($1 \le i \le n$) of each degree of stepwise hydration. The **Figure. 3** reports the Gibbs free energy change for the formation of the next hydration stages ($\Delta G_i$) of aqueous $MgCl_2$ solutions. Reactions with a negative $\Delta G$ (dark blue) occur spontaneously and with a positive $\Delta G$ (orange colour) are unfavourable and require energy input to take place. When $\Delta G$ is zero (white colour), the system has reached the maximum hydration stages. This number corresponds to the hydration number. At low concentrations, the predicted value of $h$ for $MgCl_2$(aq) is 14, which is close to the infinite dilution hydration number, 14.5, deduced from isothermal compressibility data by Marcus.[44] At in the $MgCl_2$ 2.8 mol kg$^{-1}$ solution higher concentrations $h$ decreases to 9 (**Figure.** 3).

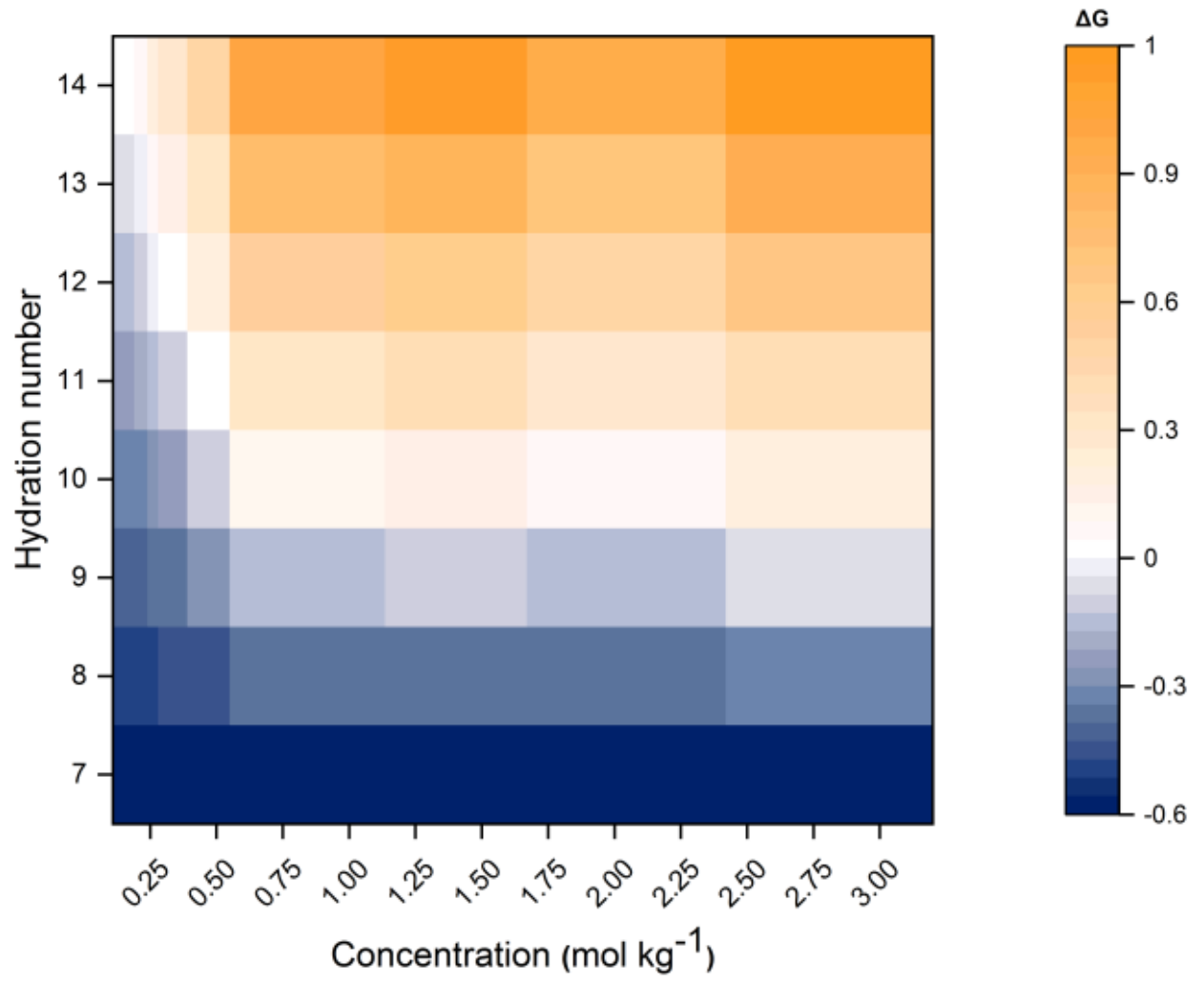

**Figure 3.** Gibbs free energy change of the stepwise hydration of $Mg^{2+}$ in aqueous $MgCl_2$ solutions ranging from 0.1 to 2.8 mol kg$^{-1}$.

We tested the same methodology on the aqueous $CaCl_2$ solutions with concentrations from 0.1 to 2.0 mol kg$^{-1}$ (**Figure. 4**). The variation of $min|\Delta\Delta G|$ as a function of the concentration and the values of $k$ are summarized in the **Figure. 4A**. The water molecules that belong to the first and second coordination shells of $Ca^{2+}$ change during the simulation, with the coordination number varying between six and seven.[45] The calcium ion is not as strongly hydrated as the magnesium ions and water exchanges between the first and second hydration shell $Ca^{2+}$ could be observed throughout the simulation compared with none of such events around $Mg^{2+}$.[46] The water molecules under the effect of chloride ions in the $CaCl_2$ solutions do not show different behaviour compared with that under the effect of a single $Cl^-$ in the pure water system. For dilute solutions (up to 0.3 mol kg$^{-1}$), the last hydration process occurs on the water molecule belonging to the $W_{22}$ or $W_{2B}$ subpopulation. For higher concentrations, this process always occurs in the $W_{21}$ subpopulations. The reason for this trend is that at low concentrations (lower than 0.5 mol kg$^{-1}$), a small number of water molecules from the $W_{22}$ subpopulation is non-bulklike, and most of the water is bulk-like. Consequently, the average result in $W_{22}$ is close to the reference $W_{BB}$ subpopulation. When the concentration is higher, most of the water molecules in the $W_{21}$ subpopulation are non-bulklike, so the average result of water in $W_{21}$ is relatively biased to $W_{1X}$. DFT calculations were also used to calculate the Gibbs free energy change of the formation of next hydration degree. There is only a small number of water molecules stably "binding" to calcium ions at the initial hydration stage ion-water cluster. The initial hydration process is:

$$Ca^{2+}(H_2O)_1 + H_2O \rightleftharpoons Ca^{2+}(H_2O)_2 \qquad K_1 = K \tag{19}$$

where the equilibrium constant parameter $K$ of $CaCl_2$ is 3.43 (Supporting Information, Table S3). The Gibbs free energy change of the formation of the next hydration stages ($\Delta G_i$) calculated with the values of $k$ and $K$ parameters are given in **Figure. 4B**. There is some disagreement between previously reported values of the hydration number of the calcium ion. Zavitsas reported a hydration number of 12.0 from freezing point depression measurements and 6.7 from boiling point elevations.[47] Moreover, the value of $h$ obtained by applying the S&R model to $CaCl_2$(aq) is the range of 4 to 8.[17] The $h$ obtained from our free energy based atomistic simulation method shows dependence on the concentration and suggest a value of 9 for dilute solutions and 4 for solution with concentrations around 0.5 mol kg$^{-1}$ or higher. In **Figure. 4B**, there is a significant increase in the hydration number for $CaCl_2$(aq) below 0.5 mol kg$^{-1}$. For these solutions, the last hydration process occurs on the water molecule belonging to the $W_{22}$ or $W_{2B}$ subpopulation (the difference of free energy is between subpopulations $W_{22}$ and $W_{BB}$). For higher concentrations, the last hydration process always occurs in the $W_{21}$ subpopulations (the difference of free energy is between subpopulations $W_{21}$ and $Cl^-$ in pure water). For concentrations lower than 0.5 mol kg$^{-1}$, a small number of water molecules from the $W_{22}$ subpopulation is non-bulklike, and most of the water is bulk-like. Consequently, the average result in $W_{22}$ is close to the reference $W_{BB}$ subpopulation. When the concentration is higher, most of the water molecules in the $W_{21}$ subpopulation are non-bulklike, so the average result of water in $W_{21}$ is relatively biased to $W_{1X}$. For $CaCl_2$(aq), the subpopulation method might be sensitive to water exchanges between the first and second hydration shell $Ca^{2+}$ occurring during the MD simulation.

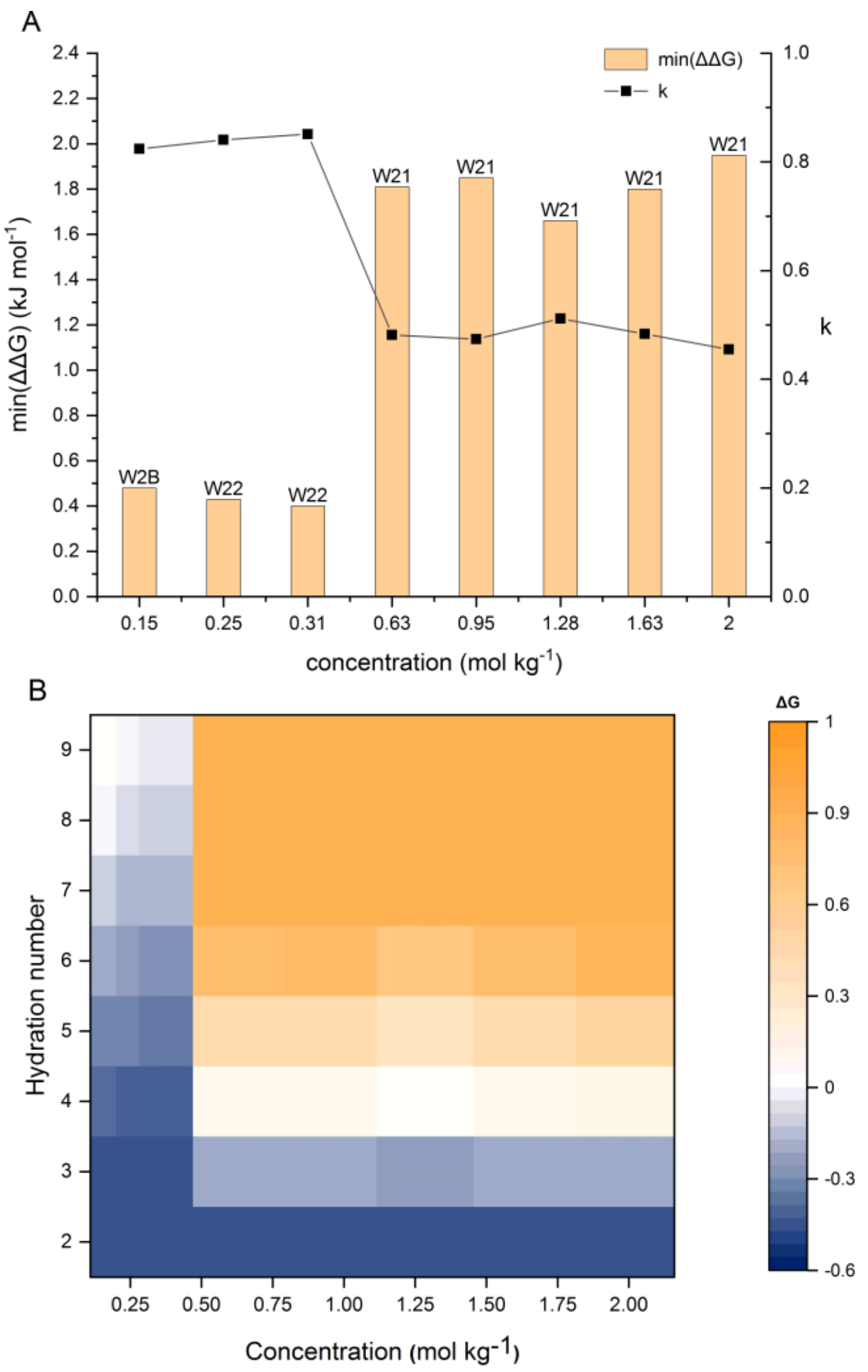

**Figure 4.** Results of $CaCl_2$(aq) from 0.1 to 2.0 mol kg$^{-1}$. **(A)** Variation of $min|\Delta\Delta G|$ and $k$ with the solution concentration. The orange bars were obtained from the difference between $\Delta G_n$ of the $Mg^{2+}$ affected subpopulations labelled in the figure and $\Delta G_s$ of the $W_{BB}$ subpopulation. **(B)** Gibbs free energy change of previous hydration stage calcium water cluster to this hydration stage calcium water cluster in $CaCl_2$ aqueous solution.

### *3.3.2 Orientational dynamics-based method*

second method to determine the hydration number parameter relies on the orientational dynamics of water molecules. Rotational motion plays a crucial role in the breaking and making of hydrogen (H) bonds (more strongly H-bonded molecules reorient more slowly) and can be computed from the first-order Legendre time correlation function:[48-50]

$$C_1(t) = \langle \vec{\mu}(0) \cdot \vec{\mu}(t) \rangle / \vec{\mu}(0)^2 \tag{20}$$

where $\vec{\mu}(0)$ and $\vec{\mu}(t)$ are the unit vectors defining the orientation of the dipole moment of $H_2O$ at times 0 and $t$, respectively. The application of this method to $MgCl_2$(aq) is reported in **Figure. 5**. The $C_1(t)$ function starts at 1 and decays asymptotically to zero because of the random and isotropic orientation of the water molecules in the solution (**Figure. 5A**). The early stage of fast loss of correlation is caused by librational motion, whereas the long term decay is due to reorientational motion and can be fit by a bi-exponential function, $a\, exp(-t/\tau_1) + b\, exp(-t/\tau_2)$.[51] The overall time associated with this process, $\tau_{reor}$, is given by the sum of fitting parameters $\tau_1 + \tau_2$.

For each subpopulation, the orientation time correlation function has been computed by tracking the dipole vectors of the water molecules belonging to that specific subpopulation. In **Figure. 5B**, we define the retardation factor as the ratio between the relaxation times of a specific subpopulation, $W_{XX}$, and bulk water, $W_{bulk}$:

$$f_{W_{XX}} = \frac{\tau_{W_{XX}}}{\tau_{W_{bulk}}} \quad (21)$$

A slow relaxation dynamic should be observed for water molecules that are in non-bulk-like subpopulations. The statistical approach used to differentiate between bulk-like and hydration water subpopulations are based on the empirical 68-95-99.7 rule. For a subpopulation $W_{XX}$ to be classified as non-bulklike the retardation factor must lie outside $3\sigma$ of the mean value of bulk-like water, that is, the value of $f_{W_{XX}}$ should be larger than the $3\sigma$ deviation. This subpopulation analysis has been conducted at each time step using four (non-overlapping) simulation blocks each lasting 5 ps. The water molecules that are in the second coordination shell of $Mg^{2+}$ (subpopulations $W_{2X}$) are outside the $2\sigma$ deviation and are classified as non-bulk-like water. A much slower relaxation dynamics compared to bulk ($5 \le f_{W_{XX}} \le 6$) is observed for water molecules that are in the first coordination shell of $Mg^{2+}$ (subpopulations $W_{1X}$). The chloride ion does not have a significant effect on the reorientation dynamics of water: subpopulations such as $W_{1B}$ with water molecules in the first coordination shell of $Cl^-$ and outside the second coordination shell of $Mg^{2+}$ have a reorientation relaxation time that is very close to that of bulk water. These results confirm the long- and short-range effects of $Mg^{2+}$ and $Cl^-$ on the reorientation water dynamics, respectively. The inset figure of **Figure. 5B** reports the distribution of water molecules per Mg-Cl ion-pairing among different subpopulations. It is also possible to evaluate the number of water molecules that are in the bulk (free water) or coordinated to $Mg^{2+}$ or $Cl^-$.

**Figure. 5C** shows the number of slow-orienting (non-bulk-like) water molecules (belonging to the subpopulations in red shadow in **Figure. 5B**) of $MgCl_2$(aq) 0.61 mol kg$^{-1}$ during the simulation time. The frequency of each hydration number during the simulation corresponds to the distribution of hydration stages, $C_i/C$, in the S&R's stepwise hydration model.[17] The different hydration degrees that exist in the solutions with concentrations ranging from 0.1 to 2.8 mol kg$^{-1}$ are reported in **Figure. 5D**.

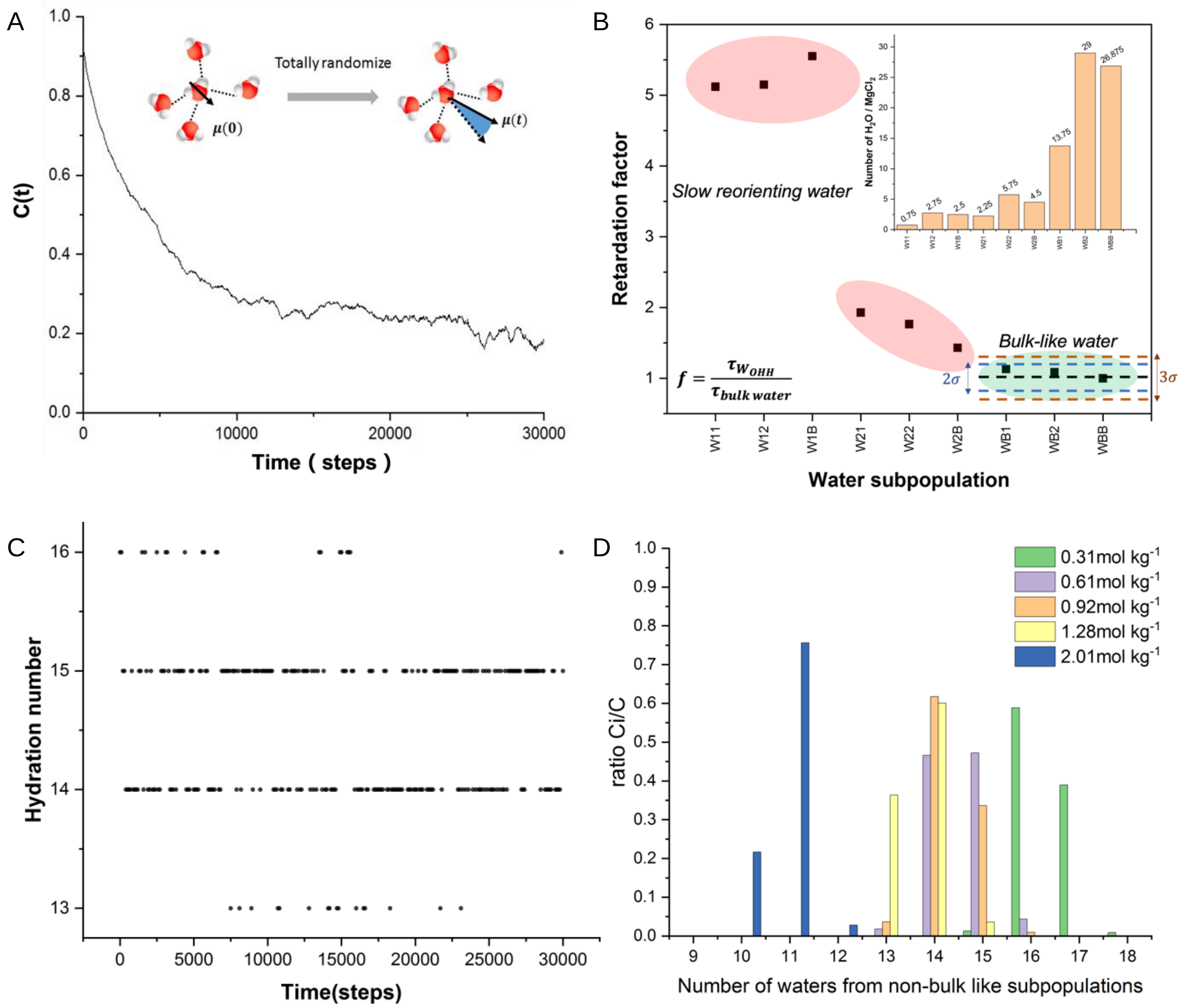

**Figure 5.** Determination of the hydration numbers of aqueous $MgCl_2$ solutions from water reorientation dynamics. **(A)** Orientational time correlation function $C_1(t)$ of the $W_{21}$ subpopulation in $MgCl_2$(aq) 0.61 mol kg$^{-1}$. **(B)** Retardation factor computed as the ratio between the reorientation relaxation time of the subpopulation and bulk water used to determine slow reorienting subpopulations in $MgCl_2$(aq) 0.61 mol kg$^{-1}$. *Inset*: Number of water molecules per Mg-Cl ion-pairing among the subpopulations. **(C)** The number of water molecules from the non-bulklike water subpopulation, corresponding to the hydration number, during each step of the MD simulation of $MgCl_2$(aq) 0.61 mol kg$^{-1}$. **(D)** Distribution of different hydration stages, $C_i/C$, existing in the $MgCl_2$ solutions.

The weighted average value of the number of waters in these non-bulklike subpopulations is the average hydration number. This figure shows a trend of the hydration number to lower values as the concentration increases.

*3.3.3 Theory and experiment comparison*

**Figure. 6** compares the hydration numbers of $MgCl_2$(aq) as a function of concentration obtained from the hydration models of S&R and Balomenos, computed from the MD simulations using the free energy and orientational dynamics methods, and determined experimentally from concentration-dependent THz dielectric relaxation (DR) spectroscopy measurements.[48] The hydration number measured from THz-DR spectroscopy corresponds to the average number of moles of water molecules per mole of dissolved salt that no longer participate in bulk-like reorientation dynamics.[48] This molecular definition of hydration number pertains to "irrotationally bound" waters (the ones tightly bound to the solute). The values from Schönert's model are not listed in **Figure. 6** because they are similar to Balomenos' model and only available up to molality of 1 mol $kg^{-1}$. The stepwise thermodynamic models cannot give hydration number parameters that change with concentration and are close to the experiment only at high concentrations. In comparison, the hydration numbers computed from atomistic properties of water molecules are concentration-dependent and close to the THz-DR experimental results.

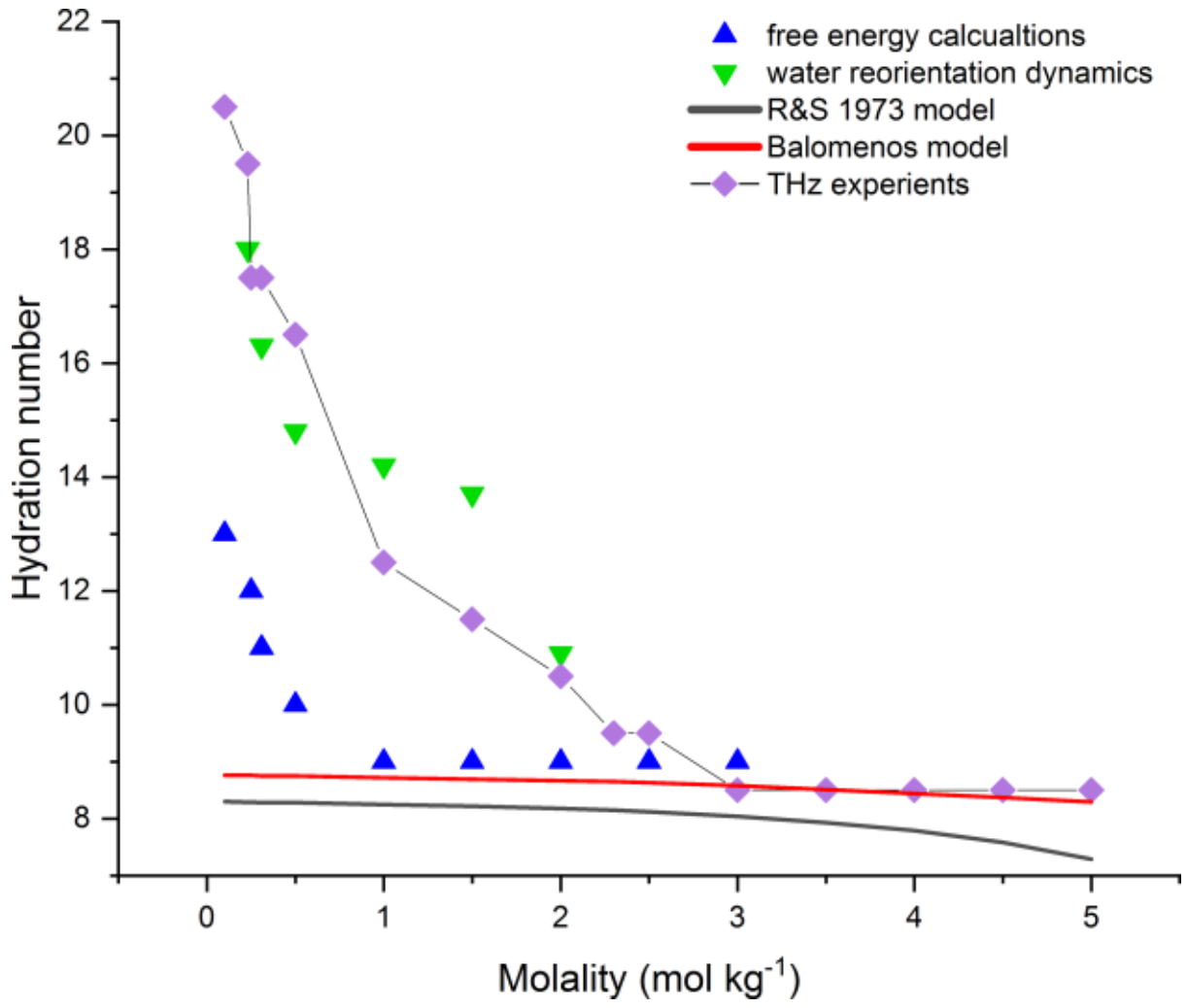

**Figure 6.** Comparison of the hydration number calculated from thermodynamic stepwise hydration models (S&R 1973 and Balomenos), the free energy- and water reorientation dynamics-based simulation methods, and THz experiments.

**Figure. 7** compares the experimental[1] and predicted values of the osmotic coefficients, $\phi$ , of $MgCl_2$(aq). We computed the osmotic coefficients from the original S&R and Balomenos models, in which the hydration number $h$ was obtained by fitting to experimental measurements, and the modified version of these two models, where $h$ has been determined in this study either from the free energy-based or water orientational dynamics-based methods. When the value of hydration number differs vastly from the original hydration number data fitting in the models, the osmotic coefficient obtained by the formula in the model will have a large gap with the results from experiments. This difference of hydration number between fitting results and computational methods results has a more noticeable impact on the osmotic coefficient as the concentration increases. For example, the osmotic coefficient from Balomenos reparametrized with the orientational dynamics-based method deviates significantly from the experimental data when concentration is higher than 1.0 mol $kg^{-1}$. The osmotic coefficient from the S&R 1973 model and the Balomenos reparameterized with the $h$ values from the free energy-based method shows a good agreement with the experiments.

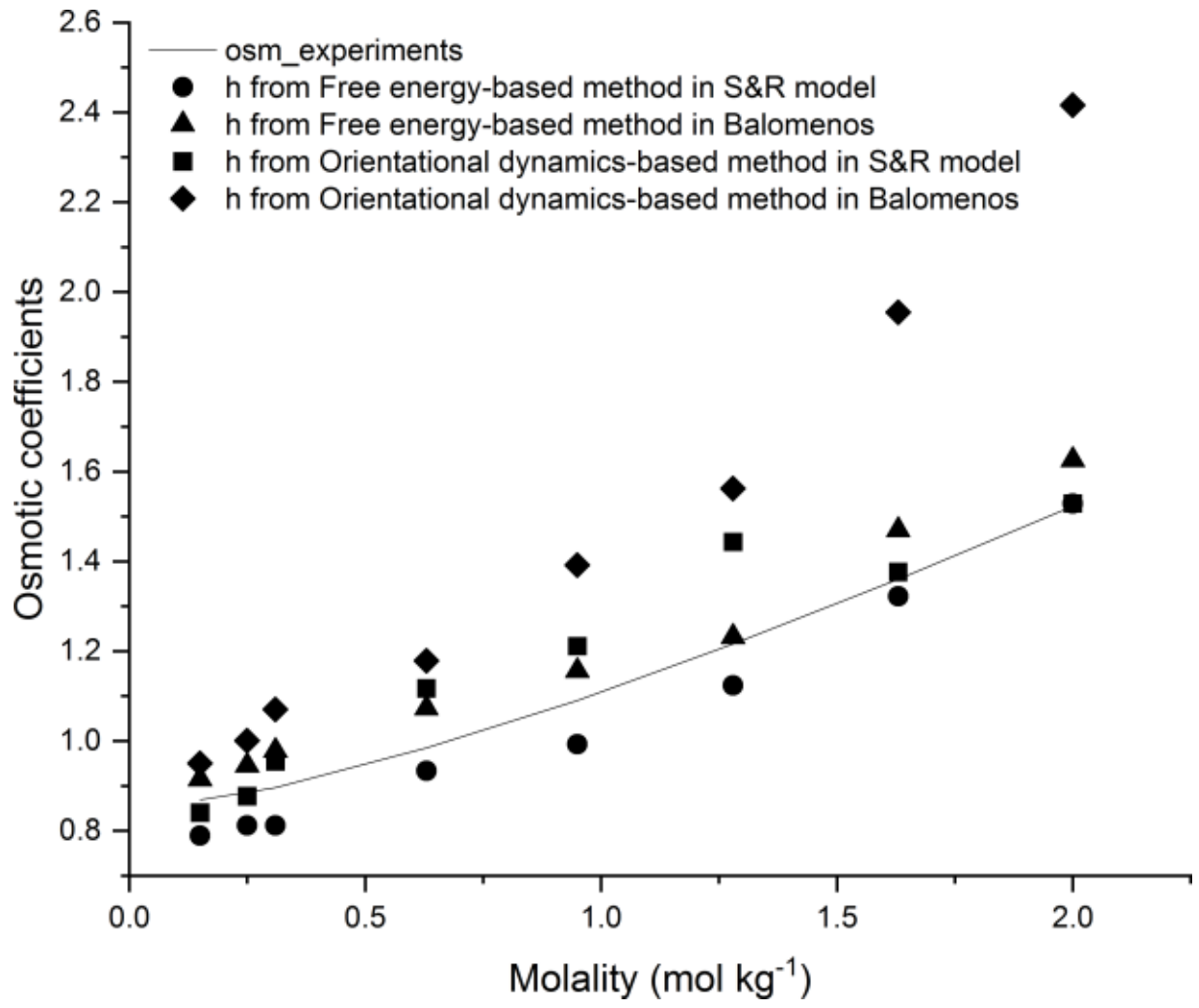

**Figure 7.** Experimental [1] and predicted osmotic coefficients, $\phi$, of $MgCl_2$ (aq). Theoretical values obtained from the original and reparameterized S&R model and Balomenos models using the free energy-based method and orientational dynamics-based method.

### 3.4 A reparameterised Stokes & Robinson model

In the previous sections, we have computed the parameters *k*, *K*, and *h*. Moreover, we have related these parameters to molecular processes occurring in the $MgCl_2$ solutions. Here, we show a simple modification of the S&R model that considers these parameters and the molecular-scale behaviour of the ions in the solution. The original S&R 1973 hydration model assumes the zeroth degree of hydration of a solute in a solution to be the solute ion itself. However, this could not be a realistic picture for most ions and, especially, strongly hydrated cations. For example, the hydrated $Mg^{2+}$ has a very stable minimum corresponding to six-fold coordination with water, $Mg(H_2O)_6^{2+}$, and the five-coordinated intermediate, $Mg(H_2O)_5^{2+}$ is inaccessible at 300 K due to the very high activation barrier between the six- and five-coordinated configurations of $Mg^{2+}$.[52] Consequently, the water exchange is drastically retarded in its first solvation shell.[45] It is possible to assume that $i$, the degree of hydration of the hydrated ion cluster, starts from 0 ($Mg^{2+}$ and six water cluster, $Mg(H_2O)_6^{2+}$) to a maximum hydration stage $n$ ($Mg^{2+}$ and $h$ number of water cluster, $Mg(H_2O)_h^{2+}$). The partial molar volumes related to the degree of hydration are given by the following expression:

$$\bar{V}_i = \bar{V}_C + (6+i)\bar{V}_W \quad (0 \le i \le n) \tag{22}$$

where $\bar{V}_C$ is the partial molar volume of solute and $\bar{V}_W$ is the partial molar volume of water. Therefore, the relationship between hydrogen stage $n$ and hydrogen number $h$ is:

$$h = 6 + n \tag{23}$$

By inserting (23) into equation (6) we obtain:

$$ln a_A = \ln[1 - c(\bar{V}_C + h\bar{V}_W)] + c[\bar{V}_C + (h-\nu)\bar{V}_W] + [\kappa^3/(24\pi N_A)]\bar{V}_W S(\kappa a) \tag{24}$$

We calculated the values of the concentration ratios $c_i/c$ using the equations:

$$c_i/c_{i-1} = (K_i/Y)a_A$$
$$c_i = c_0 \cdot (K/Y)^i \cdot k^{i(i-1)/2} a_A^i \tag{25}$$

where $lnY = c(\bar{V}_C + h\bar{V}_W - \upsilon\bar{V}_W)$. The fitting code to reproduce the hydration number, equilibrium constant parameters, distribution of different hydration degree concentrations and osmotic coefficient data from the S&R model are generated with the e04fyf Routine from NAG Library.[53]

In Eq. 25, $c_i/c$ represents the existing hydration stages in the solutions. It has been calculated by quantifying the cooperative effect of ions on the water reorientation dynamics properties of different water sub-populations. As shown in **Figure. 8**, the original and modified S&R models give different existing hydration stages of $MgCl_2$ solutions in the original S&R and modified model are different. In the modified model, because of the number of hydration stages existing in each system, the hydration number shows a wider range of values, from 7.7 to 9.3, compared with the original S&R model, from 8.6 to 7.8. Compared with the original S&R model, the osmotic coefficient results calculated from the modified model show a good fit for the experiments data, shown in **Figure. 9**. Therefore, the modified model reflects the virtual molecular-level stepwise hydration process without affecting the ability to calculate the osmotic coefficient.

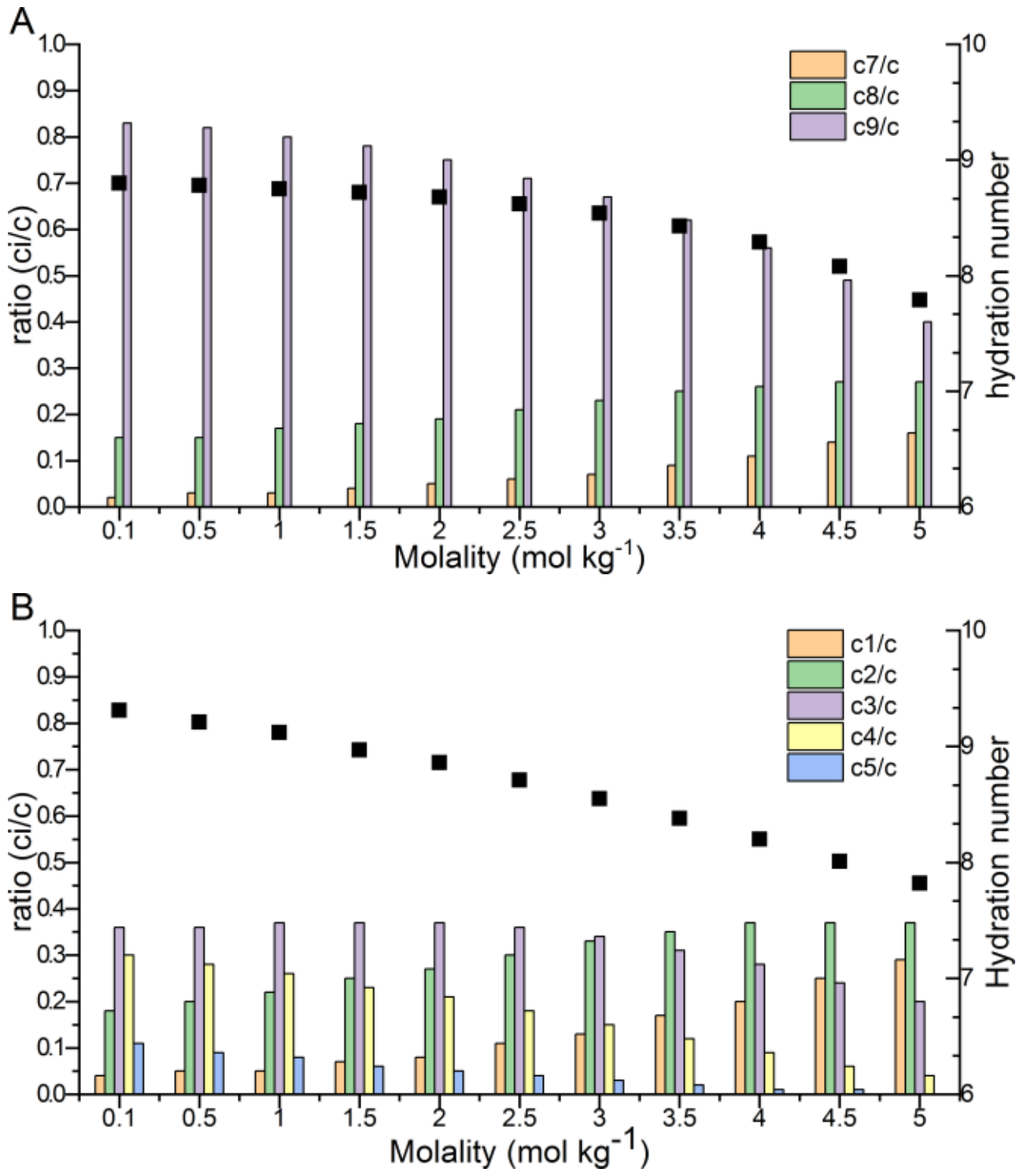

**Figure 8.** Ratio $c_i/c$ of existing hydration stages and hydration number from fitting code of the (**A**) original Stokes & Robinson 1973 and (**B**) reparameterized models.

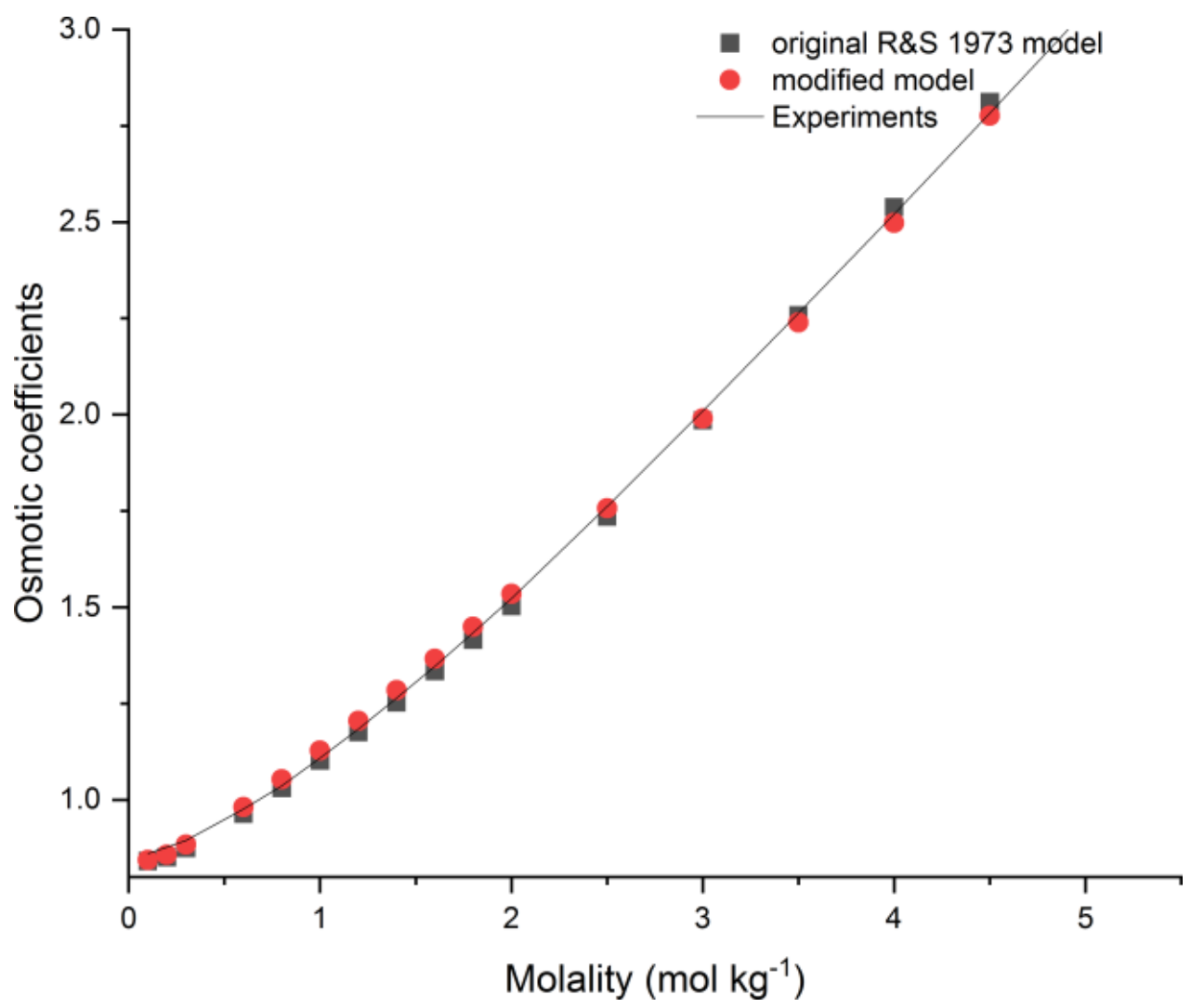

**Figure 9.** The comparison of the osmotic coefficient from experiments, original S&R 1973 model, and the modified model with the physical-meaning parameter *h*.

## 4. CONCLUSIONS

Thus far, thermodynamic models used to estimate the activity coefficients of electrolyte solutions contain empirical parameters determined by fitting the models to experimental data. In this study, we have conducted atomistic simulations to show the link between the parameters of stepwise hydration models and the molecular-level ion-hydration processes in electrolyte solutions. We have developed computational procedures to determine the concentration-dependent values of the *k*, *K* and *h* parameters used in the Stokes & Robinson (1973) and Balomenos stepwise hydration models and reparametrized such models to compute the osmotic coefficients of $MgCl_2$ solutions.

- We have implemented a simplified subpopulation water category methodology to describe the behaviour of water in the hydration shell of ions and quantify the cation-anion mixture effect. Using this classification method, we have computed the hydration free energy and reorientation dynamics of the water molecules in the subpopulations of $MgCl_2$(aq) and $CaCl_2$(aq).
- Based on the assumptions that water molecules belonging to the same subpopulation have equal ability to bind to the ion-water cluster and do not exchange between different subpopulations, we have determined the electrolyte-specific parameter *k* from the ion-affected subpopulation with the lowest absolute value of the free energy of water removal
- The equilibrium parameter, $K_1$, associated with the first degree of hydration, was computed from density functional calculations of the free energy of hydration of the hydrated ionic clusters in solution.
- The hydration number *h* was determined by considering whether the reorientation time of the water subpopulations is retarded with respect to bulk-like behaviour.
- We used the computed values of the parameters *k*, *K*, and *h* to reparametrize the hydration models of Stokes & Robinson and Balomenos and compute osmotic coefficients of aqueous $MgCl_2$ solutions as a function of concentration. The osmotic coefficients obtained from the reparameterizations of the Stokes & Robinson (1973) and Balomenos show a generally good agreement with the experiments.
- We have implemented a version of the Stokes & Robinson (1973) to describe aqueous $MgCl_2$(aq) where the highly hydrated character of $Mg^{2+}$ is part of the model. This modification leads to improvements in the values of the osmotic coefficients compared with the original model.

Our work represents an attempt to parameterize aqueous hydration models using first principle molecular-scale properties, computed from atomistic simulations, rather than fitting the models to experiments. Such an approach, if found to be generally applicable to a range of electrolyte solutions and able to incorporate a treatment of chemical equilibrium between solvent species, would remove some of the dependence on experimental thermodynamic measurements and pave the way to develop aqueous thermodynamic models capable of estimating the values of solute and solvent activities for a wide range of compositions and concentrations.

## AUTHORS INFORMATION

### Corresponding Authors

d.ditomaso@qmul.ac.uk, s.clegg@uea.ac.uk

### CRediT author statement

Xiangwen Wang: Software, Validation, Formal analysis, Investigation, Writing – original draft, Visualization. Devis Di Tommaso: Conceptualization, Methodology, Writing – review and editing, Resources, Supervision. Simon Clegg: Conceptualization, Methodology, Writing – review and editing, Resources, Supervision.

## ACKNOWLEDGMENT

XW acknowledges the QMUL Principal's Studentship for funding. DDT acknowledged the ACT programme (Accelerating CCS Technologies, Horizon2020 Project No 294766), which funded the FUNMIN project. Financial contributions made from Department for Business, Energy & Industrial Strategy (BEIS) together with extra funding from NERC and EPSRC research councils, United Kingdom, ADEME (FR), MINECO-AEI (ES). We are grateful to the UK Materials and Molecular Modelling Hub for computational resources, which is partially funded by EPSRC (EP/P020194/1). Via our membership of the UK's HEC Materials Chemistry Consortium, which is funded by EPSRC (EP/L000202), this work used the ARCHER UK National Supercomputing Service (http://www.archer.ac.uk). This research utilized Queen Mary's Apocrita HPC facility, supported by QMUL Research-IT. http://doi.org/10.5281/zenodo.438045

## DATA AVAILABILITY

Data available upon request from the authors.

# **Supplementary Material for** "Bridging atomistic simulations and thermodynamic hydration models of aqueous electrolyte solutions"

Xiangwen Wang,[a] Simon L. Clegg,[b] and Devis Di Tommaso [a]

[a] Department of Chemistry, Queen Mary University of London, Mile End Road, London, E1 4NS, UK

[b] School of Environmental Sciences, University of East Anglia, Norwich, NR4 7TJ, UK

**Corresponding Authors**

* E-mail: d.ditommaso@qmul.ac.uk, s.clegg@uea.ac.uk

## 1. Fitting with the Stokes & Robinson 1973 model

**TABLE S1.** Details of the results for $MgCl_2$ at 25°C calculated from Stokes and Robinson 1973 model (*Journal of Solution Chemistry*,**1973**, *2*, 173-191. The ion size parameter $a$, and the upper limit number of hydration sites $n$ values in the S&R model were $a$ equals to 0.4nm, $n$ equals to 4 or 5 for 1:1 electrolyte solutions or $n$ equals to 9 for 2:1 electrolyte solutions. The value of $K$ and $k$ were obtained by fitting the model to a given set of experimental activity and partial volume data. Essentially it selects a value of $K$ and finds the $k$ required to fit the most concentrated solution point. With these values, the standard deviation of the whole set of data from the equations is calculated. This is repeated with other choices of $K$ until the standard deviation is minimized. This was achieved by the e04fyf routine from NAG library. $c_i/c$ represents the existing hydration stages in the solutions.

K=12.9 k=0.904 ***a***=0.4nm n=9

| m | h | $c_0/c$ | $c_1/c$ | $c_2/c$ | $c_3/c$ | $c_4/c$ | $c_5/c$ | $c_6/c$ | $c_7/c$ | $c_8/c$ | $c_9/c$ |
|---|---|---|---|---|---|---|---|---|---|---|---|
| 0.10 | 8.80 | 3.54E-09 | 4.50E-08 | 5.16E-07 | 5.37E-06 | 5.05E-05 | 4.29E-04 | 3.31E-03 | 2.30E-02 | 1.45E-01 | 8.28E-01 |
| 0.50 | 8.78 | 6.57E-09 | 7.77E-08 | 8.32E-07 | 8.06E-06 | 7.07E-05 | 5.60E-04 | 4.02E-03 | 2.61E-02 | 1.53E-01 | 8.16E-01 |
| 1.00 | 8.75 | 1.53E-08 | 1.64E-07 | 1.60E-06 | 1.41E-05 | 1.12E-04 | 8.06E-04 | 5.25E-03 | 3.10E-02 | 1.65E-01 | 7.98E-01 |
| 1.50 | 8.72 | 3.91E-08 | 3.78E-07 | 3.30E-06 | 2.60E-05 | 1.86E-04 | 1.20E-03 | 7.05E-03 | 3.73E-02 | 1.79E-01 | 7.75E-01 |
| 2.00 | 8.68 | 1.11E-07 | 9.47E-07 | 7.34E-06 | 5.14E-05 | 3.26E-04 | 1.87E-03 | 9.73E-03 | 4.57E-02 | 1.94E-01 | 7.48E-01 |
| 2.50 | 8.62 | 3.46E-07 | 2.60E-06 | 1.76E-05 | 1.08E-04 | 6.03E-04 | 3.03E-03 | 1.38E-02 | 5.68E-02 | 2.12E-01 | 7.14E-01 |
| 3.00 | 8.54 | 1.18E-06 | 7.70E-06 | 4.53E-05 | 2.41E-04 | 1.16E-03 | 5.06E-03 | 2.00E-02 | 7.12E-02 | 2.30E-01 | 6.72E-01 |
| 3.50 | 8.43 | 4.41E-06 | 2.46E-05 | 1.24E-04 | 5.64E-04 | 2.33E-03 | 8.69E-03 | 2.94E-02 | 8.97E-02 | 2.48E-01 | 6.21E-01 |
| 4.00 | 8.29 | 1.73E-05 | 8.19E-05 | 3.50E-04 | 1.36E-03 | 4.75E-03 | 1.51E-02 | 4.32E-02 | 1.12E-01 | 2.63E-01 | 5.60E-01 |
| 4.50 | 8.08 | 7.05E-05 | 2.81E-04 | 1.01E-03 | 3.30E-03 | 9.74E-03 | 2.60E-02 | 6.29E-02 | 1.38E-01 | 2.72E-01 | 4.87E-01 |
| 5.00 | 7.79 | 2.86E-04 | 9.55E-04 | 2.89E-03 | 7.89E-03 | 1.95E-02 | 4.37E-02 | 8.86E-02 | 1.62E-01 | 2.69E-01 | 4.04E-01 |

## 2. Density functional theory calculations of the Gibbs free energy change of the initial hydration process

**TABLE S2.** Details of the results for the Gibbs free energy change of the initial hydration process of $MgCl_2$(aq) at 25°C obtained using DFT calculations. The solvation models used to conduct the solution phase calculation were the conductor-like polarizable continuum model (CPCM) and the polarizable continuum model using the integral equation formalism variant (PCM). $G_X^*$ is the total Gibbs free energy of species (X = $Mg^{2+}(H_2O)_7$, $Mg^{2+}(H_2O)_6$, or $H_2O$) in the liquid phase.

| | G*[$H_2O$] | G*[$Mg^{2+}(H_2O)_6$] | G*[$Mg^{2+}(H_2O)_7$] | ΔG* | *K* |
|---|---|---|---|---|---|
| CPCM | -47963.57144 | -413224.7137 | -461188.8784 | -0.59 | 2.72 |
| PCM | -47963.53943 | -413224.7549 | -461188.8883 | -0.59 | 2.72 |

**TABLE S3.** Details of the results for the Gibbs free energy change of the stepwise hydration processes of $CaCl_2$ at 25°C obtained using DFT calculations. The solvation model used to conduct the solution phase calculation were the conductor-like polarizable continuum model. $G_X^*$ is the total Gibbs free energy of species (X = $Ca^{2+}(H_2O)_i$, $Ca^{2+}(H_2O)_{i-1}$, or $H_2O$) in the liquid phase.

| | $G_i^*$ | $G_i^*$-$G_{i-1}^*$-$G_{H_2O}^*$ | *K* |
|---|---|---|---|
| $H_2O$ | -47963.57144 | | |
| $Ca^{2+}(H_2O)_1$ | -473090.2128 | | |
| $Ca^{2+}(H_2O)_2$ | -521054.5143 | -0.730046761 | 3.427009116 |
| $Ca^{2+}(H_2O)_3$ | -569015.3066 | 2.779186197 | 0.009197128 |
| $Ca^{2+}(H_2O)_4$ | -616977.6153 | 1.262717293 | 0.118792698 |
| $Ca^{2+}(H_2O)_5$ | -664938.2672 | 2.919537594 | 0.007257964 |
| $Ca^{2+}(H_2O)_6$ | -712901.0748 | 0.763850608 | 0.275623472 |
| $Ca^{2+}(H_2O)_7$ | -760862.8697 | 1.776507421 | 0.04992622 |
| $Ca^{2+}(H_2O)_8$ | -808823.0437 | 3.397441052 | 0.003240772 |